\documentclass[twocolumn,pre,aps,showpacs,superscriptaddress,amsmath,amssymb]{revtex4}

\usepackage{graphicx}  
\usepackage{dcolumn}   
\usepackage{bm}        
\usepackage{enumerate}
\usepackage{color}
\usepackage{psfrag}

\begin{document}

\title{Negative thermal conductivity of chains of rotors with mechanical forcing}

\author{Alessandra Iacobucci}
\affiliation{
  CEREMADE, UMR-CNRS 7534, Universit\'e de Paris Dauphine, Place du
  Mar\'echal De Lattre De Tassigny, 75775 Paris Cedex 16, France
}

\author{Fr\'ed\'eric Legoll}
\affiliation{
  Universit\'e Paris Est, Institut Navier, LAMI, Projet MICMAC ENPC - INRIA, 6 \& 8 Av. Pascal, 77455 Marne-la-Vall\'ee Cedex 2, France 
}

\author{Stefano Olla}
\affiliation{
  CEREMADE, UMR-CNRS 7534, Universit\'e de Paris Dauphine, Place du Mar\'echal De Lattre De Tassigny,
  75775 Paris Cedex 16, France
}
\affiliation{
  Universit\'e Paris Est, CERMICS, Projet MICMAC ENPC - INRIA, 6 \& 8 Av. Pascal, 77455 Marne-la-Vall\'ee Cedex 2, France 
}
\author{Gabriel Stoltz}
\affiliation{
  Universit\'e Paris Est, CERMICS, Projet MICMAC ENPC - INRIA, 6 \& 8 Av. Pascal, 77455 Marne-la-Vall\'ee Cedex 2, France
}

\date{\today}

\begin{abstract}
We consider chains of rotors subjected to both thermal and mechanical forcings,
in a nonequilibrium steady-state. 
Unusual nonlinear profiles of temperature and velocities are observed in the system.
In particular, the temperature is maximal in the center,
which is an indication of the nonlocal behavior of the system.
Despite this uncommon behavior, local equilibrium holds for long enough chains.
Our numerical results also show that, when the mechanical forcing is strong enough,
the energy current can be increased by an inverse temperature gradient.
This counterintuitive result again reveals the complexity of nonequilibrium states.
\end{abstract}


\maketitle

%
%

\section{Introduction}

Thermodynamic properties of non-equilibrium stationary states are
poorly understood. They are usually characterized by currents of
conserved quantities, such as energy, flowing through the system.
When stationary states are \emph{close} to equilibrium states,
linear response theory is effective and explains common macroscopic
phenomena like Fourier's law: In a system 
in contact with two thermostats at different temperatures, the heat flux 
is proportional to the temperature gradient (as long as the relative
difference between the two temperatures is small).

In contrast, there is no general theory to describe systems 
in a stationary state {\em far} from equilibrium, and the
corresponding macroscopic properties seem to depend
on the specific details of the dynamics.

In this article, we investigate numerically the energy transport
properties of a simple one-dimensional system, a chain of $N$~rotors,
in a stationary state far from equilibrium. 
Many studies considered one-dimensional chains of oscillators subjected to a temperature
gradient~\cite{sll,Dhar}.
Here we consider both thermal and mechanical forcings, obtained as follows: 
The leftmost rotor is attached to a wall and put in contact with a Langevin 
thermostat at temperature $T_{\rm L}$, while the rightmost rotor is subjected to 
a constant external force $F$ and put in contact with another Langevin thermostat at
temperature $T_{\rm R}$.  

What we observe in our numerical experiments is that {\em the combined effect
of these two generalized forces can reduce the current instead of
increasing it}. This counterintuitive effect 
is observed for large mechanical forcings $F$, 
when $T_{\rm R}$ is increased while $T_{\rm L}$ remains fixed
(see Figure~\ref{fig:TLfixed} below). The mechanical forcing induces a
negative current (from the right to the left). When $T_{\rm R}$ is increased
while $T_{\rm L}$ remains fixed, one would naively expect that the
(negative) thermal forcing is also larger, and thus that the negative
current should be (in absolute value) larger. In contrast to this
expectation, we observe that, in this case, the current
is {\em reduced}.
This strange effect does not appear if, instead, $T_{\rm L}$ is lowered
and $T_{\rm R}$ is fixed (see Figure~\ref{fig:TRfixed} below), in which
case the current indeed becomes larger in absolute value. 

We are unable to provide explanations to the above described phenomena.
We believe that such behaviors show the complexity of non-equilibrium
stationary states far from equilibrium, and also suggest that
Fourier's law is only valid close to equilibrium. A naive
extension of the definition of thermal conductivity to genuinely non-equilibrium settings
can give negative values to this quantity (whence the somehow
provocative title of this article).

In the sequel of this article, we first describe the system we work
with, and the numerical integrator we have used (see
Sec.~\ref{sec:description}). We next turn to
studying various properties of the system. In particular, we numerically
check that local equilibrium holds for systems large enough, despite the fact that, globally, the
system is out of equilibrium (see Sec.~\ref{sec:local}). In
Sec.~\ref{sec:current}, we study
how the current depends on 
the magnitude of the mechanical force and on the temperatures that are
imposed on both ends of the chain. All these numerical studies are
performed for chains of increasing lengths.

\section{Description of the system}
\label{sec:description}

The configuration 
of the system is described by the positions (angles) $q = (q_1,\dots,q_N)$ 
of the rotors, which belong to the one-dimensional
torus $2\pi\mathbb{T}$, as well as their associated (angular)
momenta $p = (p_1,\dots,p_N)$. The masses of the particles are set to~1 for 
simplicity. The Hamiltonian of the system is 
\begin{equation}
  \label{eq:Hamiltonian}
  H(q,p) = \sum_{i=1}^N \left[ \frac {p_i^2}2 + (1 - \cos r_i)\right],
\end{equation}
where we have set $r_i = q_i - q_{i-1}$ for $i \geq 2$ and $r_1 = q_1$.

We consider a system with free boundary conditions on the right end, whose
evolution equations read:
\begin{equation}
\label{eq:dynamics}
\left\{ \begin{aligned}
dq_i & = p_i \, dt, \\
dp_i & = \Big( \sin(q_{i+1}-q_i) - \sin(q_i-q_{i-1}) \Big) dt,\ i\neq
1, N, \\
dp_1 & = \Big( \sin(q_2-q_1) - \sin(q_1) \Big) dt \\
& \qquad\qquad - \gamma p_1 \, dt + \sqrt{2\gamma T_{\rm L}} \, dW^1_t,\\
dp_N & = \Big(F- \sin(q_N-q_{N-1}) \Big) dt \\
& \qquad\qquad - \gamma p_N \, dt + \sqrt{2\gamma T_{\rm R}} \, dW^N_t,\\
\end{aligned} \right.
\end{equation}
where $W^1_t$ and $W^N_t$ are independent standard Wiener processes, and
$\gamma>0$ determines the strength of the coupling to the thermostat. In
the sequel, we work with $\gamma=1$.
Note that the external constant force $F$ is non-gradient since it does not
derive from a periodic potential.

We checked the robustness of the results we describe below
with respect to the choice of boundary conditions. We indeed also considered
fixed boundary conditions on the right end (this amounts to adding an
extra force $-\sin(q_{N})$ to the last atom).
In particular, we checked that our counterintuitive
results on the behavior of the thermal current as a function of the strength
of the nongradient force are still observed with
these boundary conditions.

\subsection{Equilibrium and nonequilibrium states}

If $F=0$ and $T_{\rm L} = T_{\rm R} = T$, the system is in
\emph{equilibrium}, and the unique stationary measure is given by the
Gibbs measure at temperature $T$ associated with the
Hamiltonian~\eqref{eq:Hamiltonian}. 
When $T_{\rm L} \neq T_{\rm R}$ with $F=0$, the properties of the
non-equilibrium stationary state have
been studied numerically by various 
authors~\cite{Giardina:2000p10893,GendelmanSavin,YangHu,GendelmanSavinReply}. 
The thermal conductivity of the system, defined as the stationary
energy current multiplied by the size of the system and
divided by the temperature difference~\cite{blr}, has a finite limit
for large system sizes, even though
the rotor chain is a momentum conserving one-dimensional  
system~\cite{bborev,bbo2}. Besides,
as the average temperature $T$ increases above the value~0.5, the thermal
conductivity decreases dramatically~\cite{Giardina:2000p10893}. 

If $F\neq 0$, the system is out-of-equilibrium
even if $T_{\rm L} = T_{\rm R}$ (recall indeed that $F$ is non-gradient). The force, in the
stationary state, induces an energy current towards the left. 
The stationary state cannot be computed explicitly and, if $F$ is
large, linear response theory cannot be used to obtain information about the
conductivity of the system.   

If $T_{\rm L} < T_{\rm R}$, there are two mechanisms that {\em separately}
generate an energy current towards the left of the system: The
mechanical force $F$ and the thermal force given by the
temperature gradient. 
It seems however difficult to separate the contributions of each mechanism.
The numerical experiments reported below show that these two mechanisms are not
necessarily additive, and that one mechanism may reduce the effect of
the other one, leading to counterintuitive results.

\subsection{Numerical integration}

The numerical integration of~\eqref{eq:dynamics} is
performed using a splitting strategy 
where the Hamiltonian part of the evolution is integrated with the Verlet
scheme~\cite{Verlet}.
The fluctuation-dissipation parts, with the additional
non-gradient force, are Ornstein-Uhlenbeck processes and can thus be
integrated analytically. We have thus used the following algorithm:
\begin{eqnarray*}
\widetilde{p}^n_1 &=& \alpha p^n_1 + \sigma_{\rm L} G^n_1,
\\
\widetilde{p}^n_N &=& F + \alpha (p^n_N-F) + \sigma_{\rm R} G^n_N,
\\
\widetilde{p}^n_i &=& p^n_i, \quad i \neq 1,N,
\\
p^{n+1/2}_i &=& \widetilde{p}^n_i - 
\frac{\Delta t}{2} \frac{\partial H}{\partial q_i}(q^n,\widetilde{p}^n),
\\
q^{n+1}_i &=& q^n_i + \Delta t \, p^{n+1/2}_i,
\\
p^{n+1}_i &=& p^{n+1/2}_i - 
\frac{\Delta t}{2} \frac{\partial H}{\partial q_i}(q^{n+1},p^{n+1/2}),
\end{eqnarray*}
where $\alpha = \exp(-\gamma \Delta t)$, 
$\displaystyle \sigma_{\rm L} = \sqrt{(1-\alpha^2) T_{\rm L}}$,
$\displaystyle \sigma_{\rm R} = \sqrt{(1-\alpha^2) T_{\rm R}}$, and $H$
is given by~\eqref{eq:Hamiltonian}. In turn,
$G^n_1$ and $G^n_N$ are independent normal Gaussian random
variables. Recall also that the friction parameter $\gamma$ is set to~1.
The three first lines of the above algorithm consist in
exactly integrating the Ornstein-Uhlenbeck processes on $p_1$ and $p_N$,
whereas the three last lines are based on the standard Verlet algorithm. 

The time-step $\Delta t = 0.05$ ensures that the energy conservation in the Verlet scheme is 
accurate enough. While there might be some time-step bias in the value of the currents,
the qualitative conclusions are robust with respect to the choice of the time-step.

\section{Properties of the nonequilibrium system}

This section is organized as follows. First, we discuss the existence of
a stationary measure for the dynamics~\eqref{eq:dynamics}. Under the
assumption that such a stationary measure exists, we establish some
relations that are consistent with physical intuition. We next point out
that this system shows some very surprising
features. For instance, the temperature profile is non-monotonic and a
maximum is observed in the center of the system, while the velocity
profiles are very nonlinear. Despite these nonlocal features, we show
that local equilibrium holds. We finally turn to investigating the
dependence of the stationary energy current on $F$, $T_{\rm L}$ and
$T_{\rm R}$. 

\subsection{Stationary measure}

We believe that there exists a unique smooth stationary measure
for the dynamics~\eqref{eq:dynamics}. However, 
as far as we know, there is no rigorous result in
this direction for rotor chains, even in the case $F = 0$. 
Indeed, the standard techniques (see for
instance~\cite{ReyBellet,carmona}) used to prove existence and 
uniqueness of an invariant measure for chains of oscillators
under thermal forcing do not apply here.

\begin{figure}[htbp]
\begin{center}
\includegraphics[angle=270,width=8.cm]{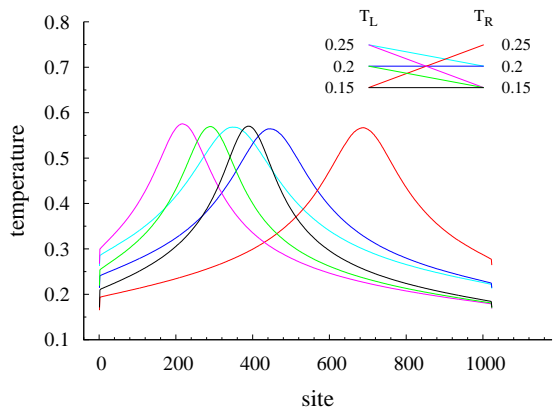}
\caption{\label{fig:temperature}
Kinetic temperature profiles for chains of length $N = 1024$, with $F =
1.6$ and different temperature gradients.
}
\end{center}
\end{figure}

A possible pathology for rotor chains is that the (internal) energy concentrates 
locally on one or several rotors, which rotate faster and faster. Since the
interaction forces are bounded, it may not be possible to prevent this 
fast rotation. In practice, we have not observed such catastrophes in the 
parameter regime we considered, but the kinetic temperature profiles presented
in Figure~\ref{fig:temperature}
(obtained from the variance of the momenta, with the previously mentioned
caveat on the interpretation of this quantity) are quite unexpected and 
show that the internal energy tends to be larger in the middle of the chain.
This picture also allows to understand what happens when the imposed
temperatures at the right and left ends change:
The maximal temperature in the chain is almost unchanged, but the position of the maximum is
displaced. 
This shows that the linear response correction to the stationary measure is necessarily 
\emph{nonlocal}. Such nonlocal effects were already observed in nonequilibrium
exclusion processes~\cite{DLS02a,DLS02b}.

Some interesting relations can nonetheless be obtained under the assumption
that the stationary state exists.
We denote by $\langle\cdot\rangle$ the expectation with respect to the stationary
measure. 
First, a constant profile of force settles down in the bulk. Taking expectations 
in~\eqref{eq:dynamics} indeed gives
\[
\begin{aligned}
\langle \sin r_{i+1}\rangle  & = \langle \sin r_i\rangle, 
\qquad \qquad \qquad i \neq 1,N,\\
\langle \sin r_2\rangle  & = \langle \sin r_1\rangle  + \gamma \langle
p_1\rangle,
\\
\langle \sin r_N\rangle & = F - \gamma \langle p_N\rangle.
\end{aligned}
\]
This leads to the following profile:
$F_i := \langle \sin r_i\rangle  = F - \gamma \langle p_N\rangle$ for all $i \ge 2$,
while $F_1 := \langle \sin r_1\rangle = F - \gamma (\langle p_N\rangle + \langle p_1\rangle )$.

The balance between the average work done by the force and
the energy dissipated by the thermostats is given by
\begin{equation}
\label{eq:energy_balance}
0 = F \langle p_N\rangle  + \gamma(T_{\rm L} - \langle p_1^2\rangle ) + \gamma (T_{\rm R} -
\langle p_N^2\rangle ), 
\end{equation}
as can be seen by noticing that the average variation of the total
energy~$H$ is zero. 
Moreover, the entropy production inequality (obtained by computing the
variations of the relative entropy with respect to the invariant
measure, see e.g.~\cite{bo}) 
gives
$$
  T_{\rm L}^{-1}(T_{\rm L} - \langle p_1^2\rangle ) + T_{\rm R}^{-1}(T_{\rm R} - 
  \langle p_N^2\rangle ) \le 0.
$$
In the case $T_{\rm L} = T_{\rm R} = T$, this relation, 
combined with~\eqref{eq:energy_balance}, yields $F \langle
p_N\rangle \ge 0$. Therefore the stationary momentum
on the right end has the same sign as the driving force, as
expected.

\begin{figure}[htbp]
\begin{center}
\includegraphics[angle=270,width=8.cm]{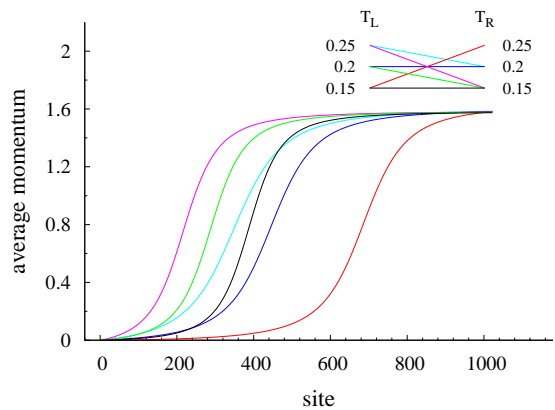}
\caption{\label{fig:velocities}
Average momenta for chains of length $N = 1024$, with $F = 1.6$.
}
\end{center}
\end{figure}

Figure~\ref{fig:velocities} shows that the momentum profile
is not linear, and that its derivative is maximal where, according to
Figure~\ref{fig:temperature}, temperature is maximal. We also observe on 
Figure~\ref{fig:thermo_limit} (top) that the profile seems to become
steeper in the thermodynamic limit.

\subsection{Local equilibrium and thermodynamic limit}
\label{sec:local}

A very interesting question is whether nonequilibrium systems
are {\em locally} close
 to equilibrium. This issue was considered in~\cite{Mai_et_al}
for systems subjected to thermal forcings only.
We check the local equilibrium assumption in three steps.

\begin{enumerate}[(i)]
\item We study the agreement between the local kinetic temperature
(defined as the variance of the velocities) and the local potential
temperature. The latter is obtained as follows. First, we numerically
precompute the function 
$$
g : 
T \mapsto \frac{\displaystyle 
\int_0^{2\pi} V(r) \exp(-V(r)/T) \, dr}{\displaystyle 
\int_0^{2\pi} \exp(-V(r)/T) \, dr}
$$
which, to a given temperature, associates the canonical average of the
potential energy $V(r) = 1 - \cos r$ of one bond. The local potential
temperature at bond $i$ is then defined as the value $T_i$ such that
$g(T_i)$ is equal to the time average of the potential energy of
the bond $r_i$ along the trajectory defined by~\eqref{eq:dynamics}. 
The results presented in Figure~\ref{fig:thermo_limit} (bottom) show that 
the two local temperatures are quite different for small systems, but are identical
for larger ones. Besides, as the length of the system increases, the profiles become more 
symmetric.
\begin{figure}[htbp]
\begin{center}
\includegraphics[angle=270,width=8.cm]{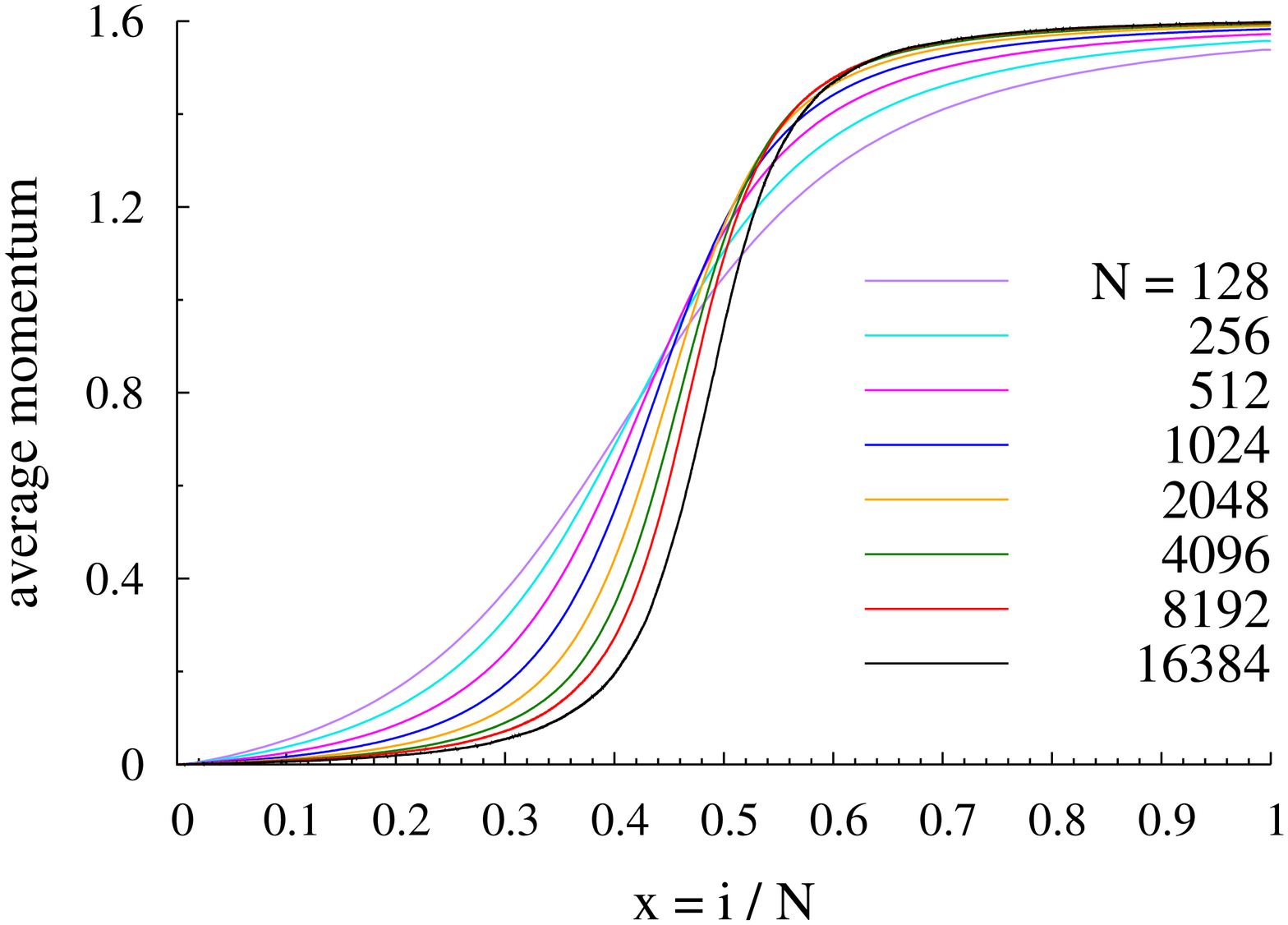}
\includegraphics[angle=270,width=8.cm]{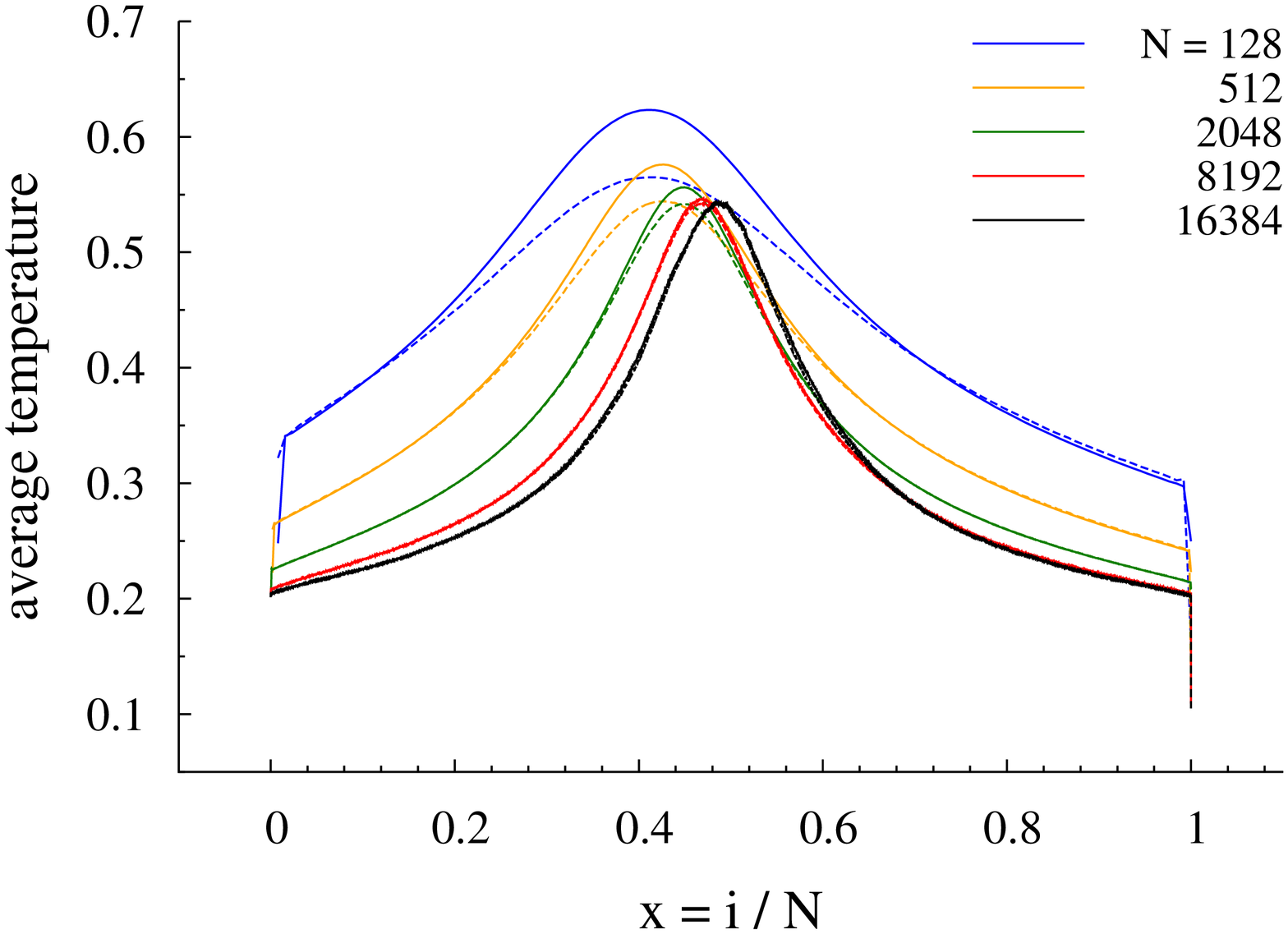}
\caption{\label{fig:thermo_limit}
Rescaled profiles for systems of increasing size
$N = 2^k$ with $k = 7,\dots,14$: the $x$ variable 
is the site index~$i$ divided by $N$. The value of the nongradient force is
$F=1.6$ and $T_{\rm L} = T_{\rm R} = 0.2$.
Top: momenta. Bottom: kinetic (solid lines) and potential (dashed lines)
temperatures.
}
\end{center}
\end{figure}

\item We check that the individual distributions of~$p$ and~$r$ 
are in accordance with a local Gibbs equilibrium. To this end, we build
the histograms of the momenta and distances at the site~$i_{\rm max}$
where the local temperature is maximal (since this is the location where the
disagreement between the local kinetic and potential temperatures 
is the strongest). The results presented in Figure~\ref{fig:histograms} 
show that the empirical distributions of $p$ and~$r$ at the site~$i_{\rm
  max}$ are in excellent agreement with the Gibbs distributions with the
same parameters (average velocity $\overline{p}_i$, temperature
$T_i$), namely 
$$
Z_{\rm kin}^{-1} \exp \left[ -(p - \overline{p}_i)^2/(2T_i) \right] \, dp
$$
and
$$
Z_{\rm pot}^{-1} \exp \left[ -V(r)/T_i \right] \, dr,
$$
except for the smallest systems (say, $N \leq 512$). 

\begin{figure}[htbp]
\begin{center}
\psfrag{momenta}{momentum}
\psfrag{distances}{distance}
\includegraphics[width=6.cm,angle=-90]{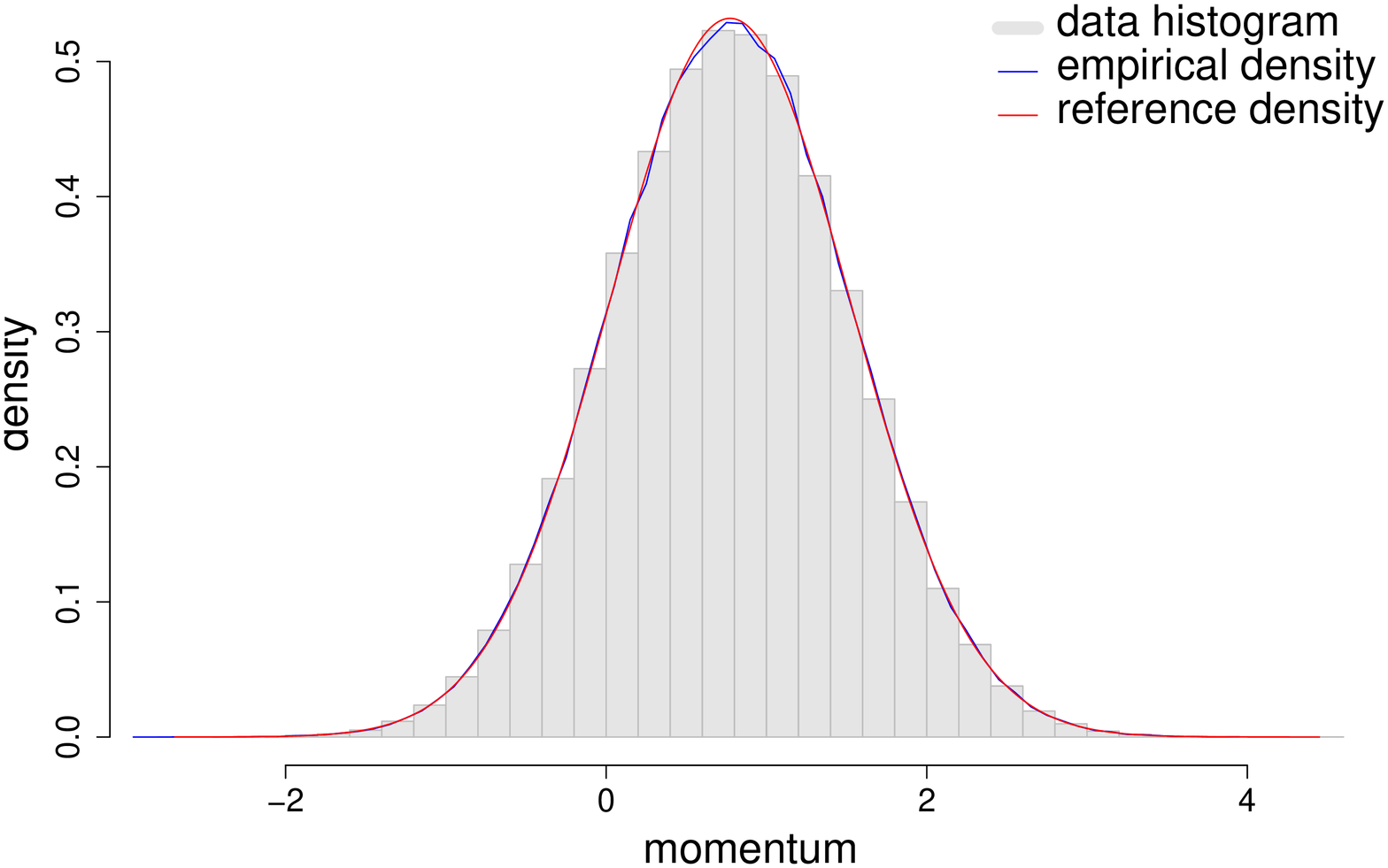}
\includegraphics[width=7.cm]{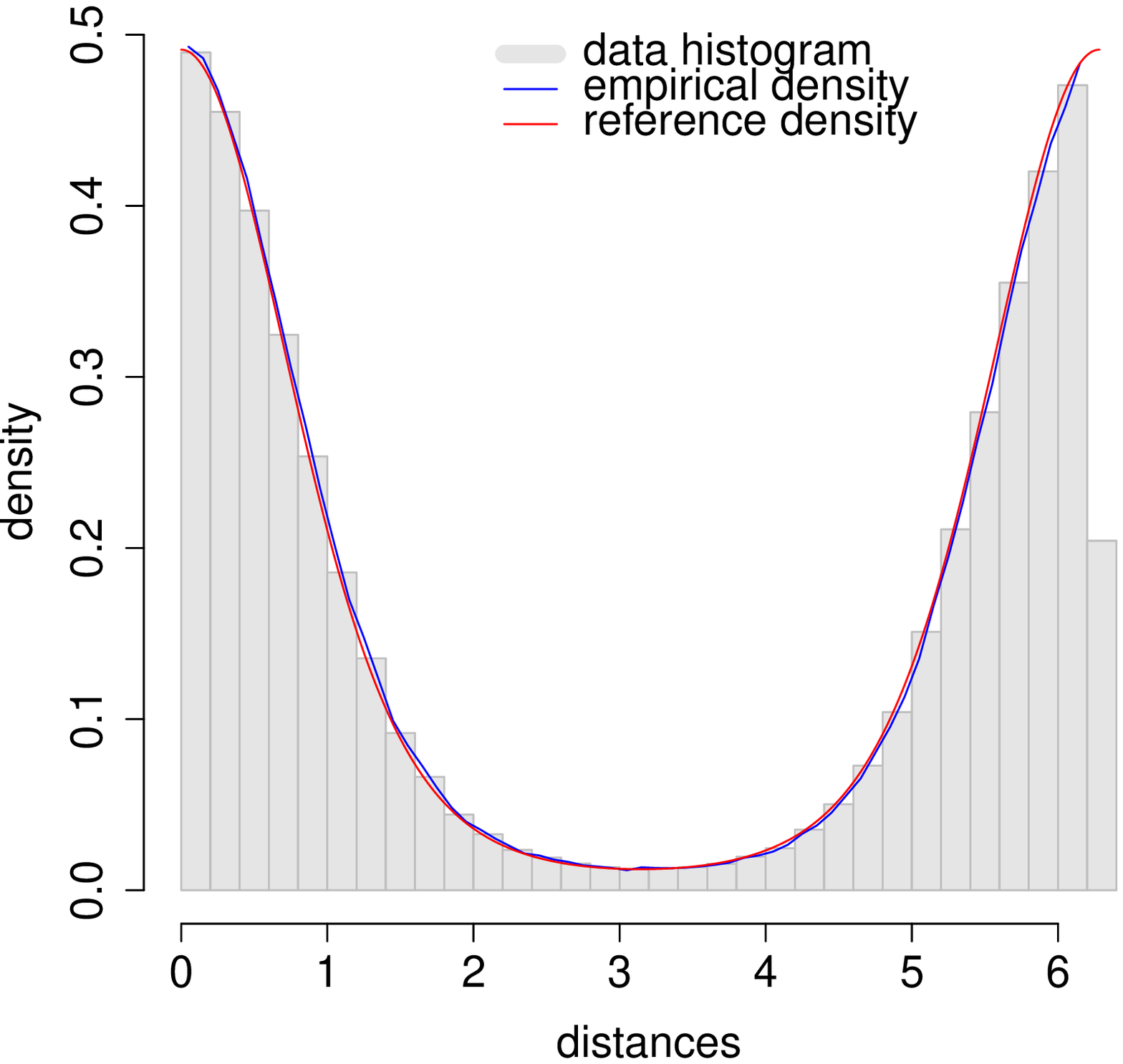}
\caption{\label{fig:histograms}
Top: Empirical distribution of momenta at the site $i_{\rm max}$ 
(where the temperature is maximal), and comparison with the local
Gibbs equilibrium with the same average and variance. 
Bottom: Empirical distribution of the distances at bond $i_{\rm max}$ and
comparison with the local Gibbs equilibrium with the same average
energy. 
Both plots correspond to a chain of length $N=1024$, with
$F=1.6$ and $T_{\rm L} = T_{\rm R} = 0.2$.
}
\end{center}
\end{figure}

\item We check that momenta and distances are independent. To this end, 
we compare the joint law $\psi = \psi(r_{i_{\rm max}}, p_{i_{\rm max}})$ 
of $(r_{i_{\rm max}}, p_{i_{\rm max}})$
and the product law obtained from the tensor product of the individual 
distributions of these two variables (denoted respectively
by $\overline{\psi}_r(r_{i_{\rm max}})$ and $\overline{\psi}_p(p_{i_{\rm max}})$). 
More precisely, for a given number $n$ of sample points (obtained by
subsampling a long trajectory every $10^4$~steps), we check that the 
distance
\begin{equation}
\label{eq:error}
\delta_n = \int_{[0,2\pi] \times \mathbb{R}} 
\left|\psi^n(r,p)-\overline{\psi}^n_r(r)
\overline{\psi}^n_p(p)\right| 
\, dr \, dp
\end{equation}
between these two distributions indeed decreases as the inverse square-root 
of the number of configurations used to build the histograms. 
Again, this is true for systems large
enough. Figure~\ref{fig:joint_error} shows that $\delta_n \sim
n^{-1/2}$. 

\begin{figure}[htbp]
\begin{center}
\psfrag{chain length}{$\log_2 n$}
\psfrag{joint error}{$\log_2 \delta_n$}
\includegraphics[width=8.cm]{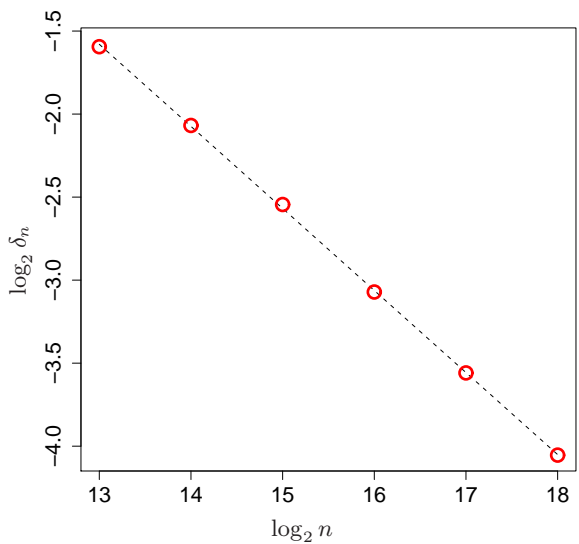}
\caption{\label{fig:joint_error}
Decrease of the error~$\delta_n$ defined by~\eqref{eq:error} as a function
of the number of sample points~$n$. Estimated rate of decrease:
$\delta_n \simeq 28.7 \times n^{-0.494}$. 
}
\end{center}
\end{figure}
\end{enumerate}

\subsection{Behavior of the energy current}
\label{sec:current}

\begin{figure}[htbp]
\begin{center}
\includegraphics[angle=270,width=8.cm]{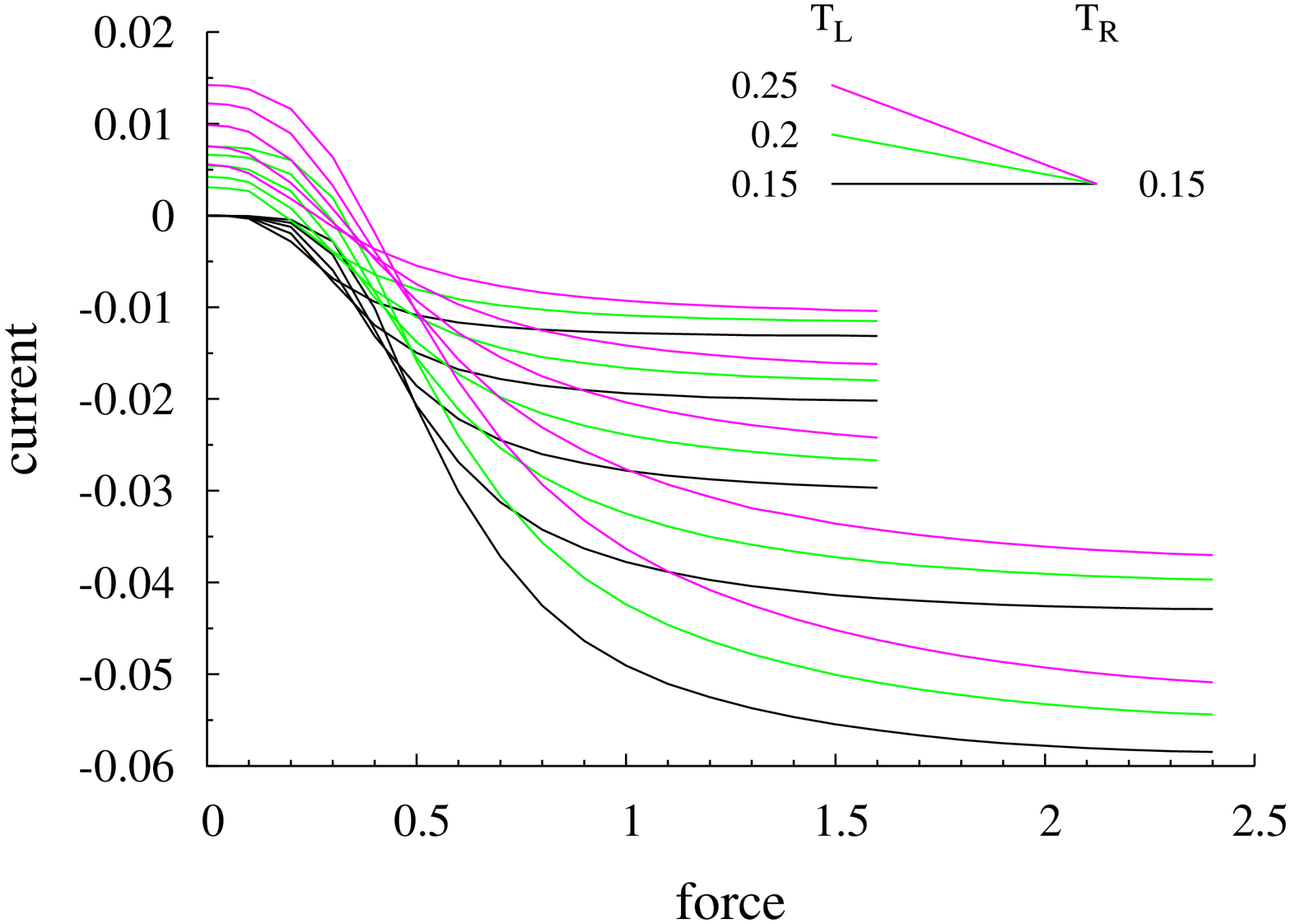}
\includegraphics[angle=270,width=8.cm]{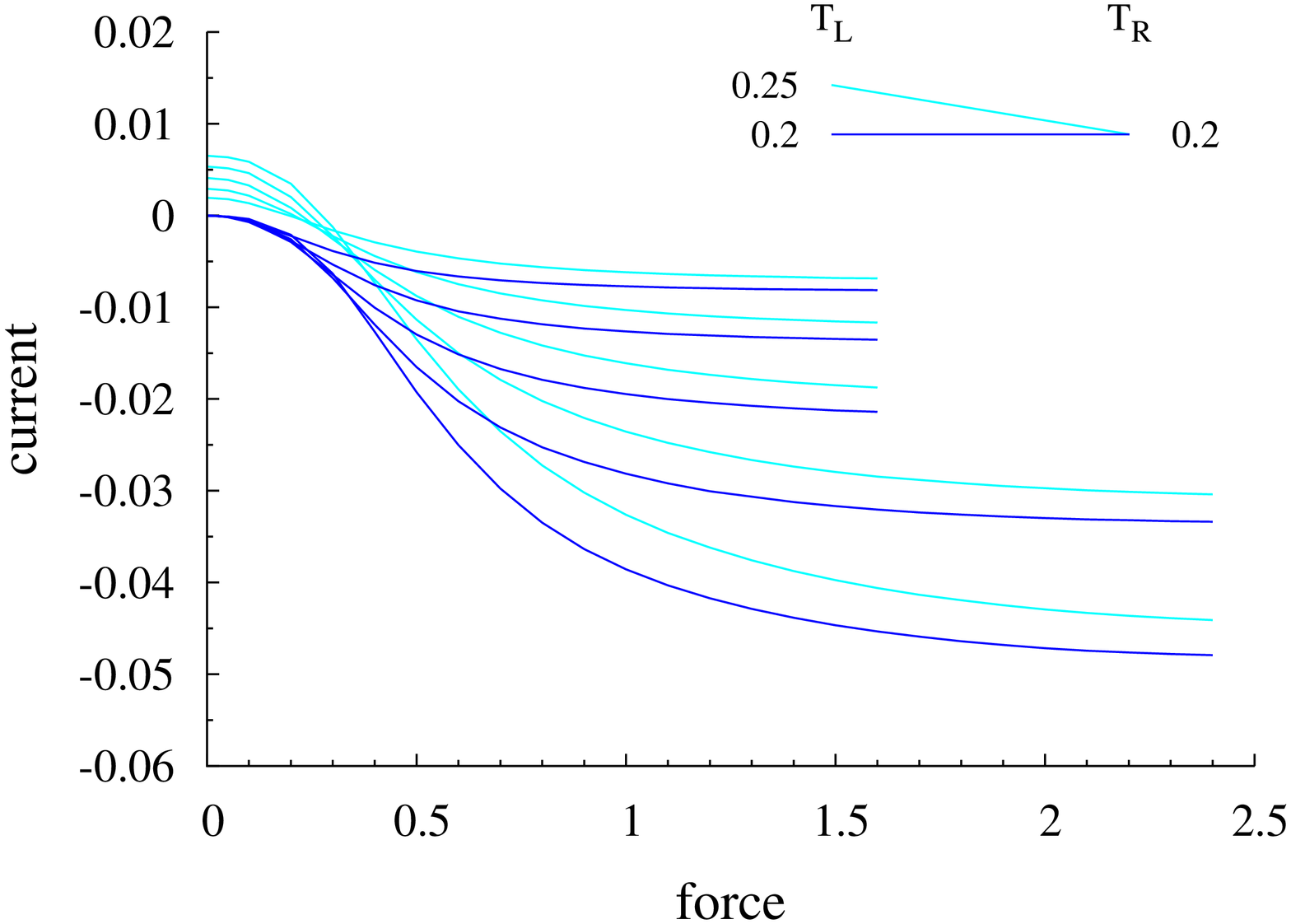}
\caption{\label{fig:TRfixed}
Comparison of the currents with fixed temperature on the right end
and increasing temperatures on the left end. 
From top to bottom: decreasing system sizes $N=2048,1024,512,256,128$
(the ordering is the same for all situations considered; for the longest
systems, we have considered forces $0 \leq F \leq 1.6$; for the shortest
ones, we have considered the range $F \in [0;2.4]$).
}
\end{center}
\end{figure}
\begin{figure}[htbp]
\begin{center}
\includegraphics[angle=270,width=8.cm]{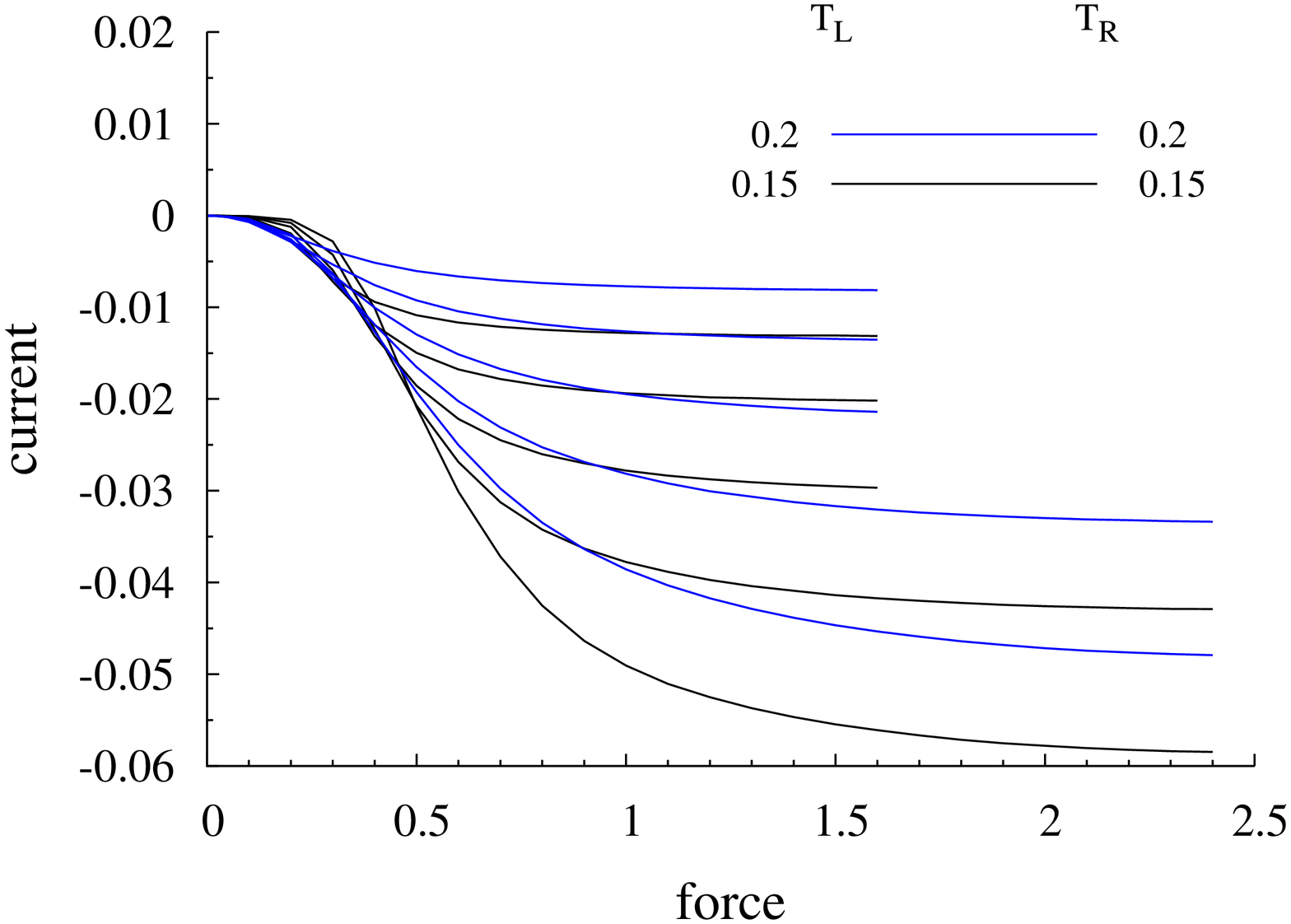}
\includegraphics[angle=270,width=8.cm]{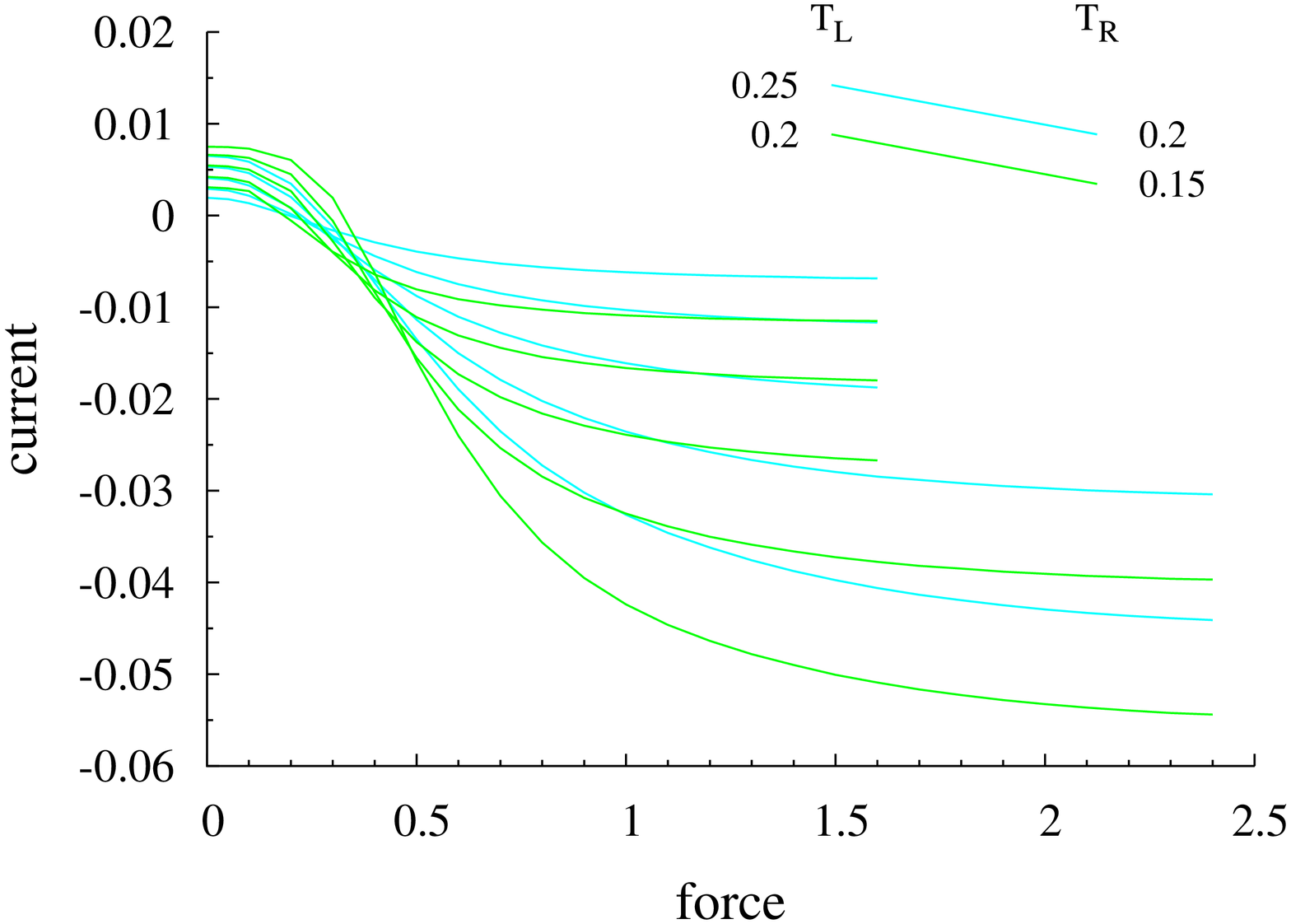}
\caption{\label{fig:dTfixed}
Comparison of the currents with fixed temperature difference.
From top to bottom: decreasing system sizes $N=2048,1024,512,256,128$
(the ordering is the same for all situations considered). 
}
\end{center}
\end{figure}
\begin{figure}[htbp]
\begin{center}
\includegraphics[angle=270,width=8.cm]{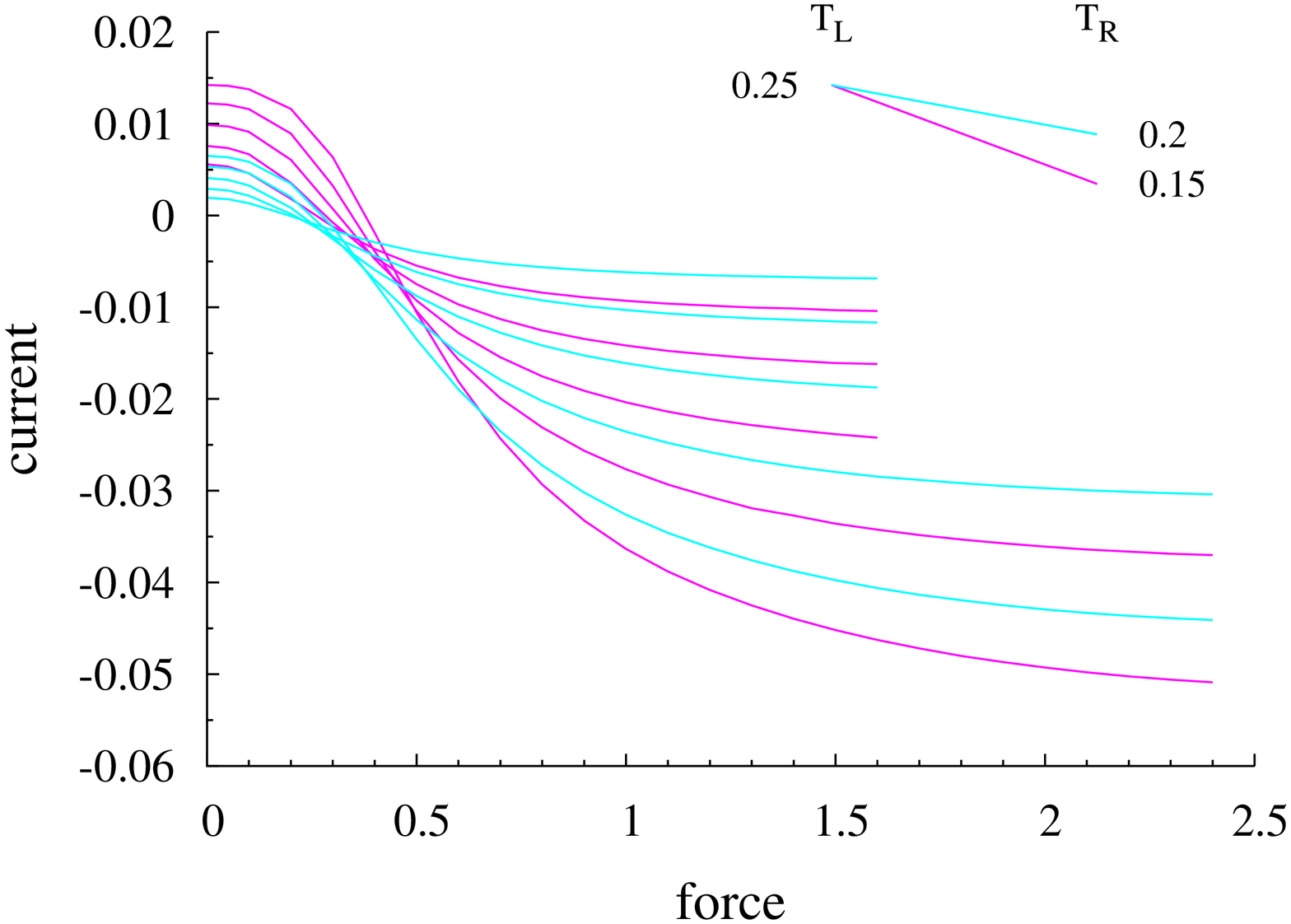}
\includegraphics[angle=270,width=8.cm]{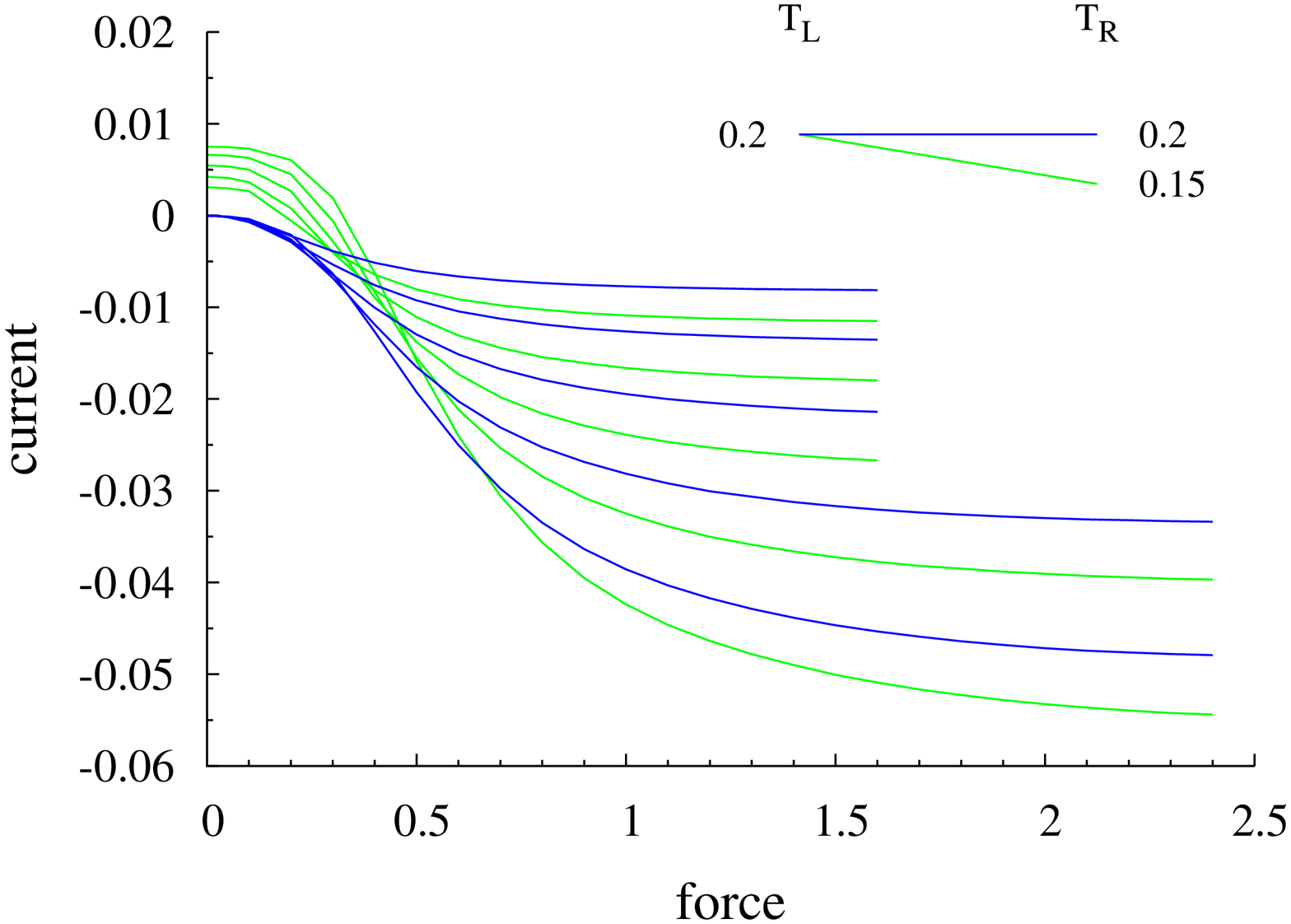}
\includegraphics[angle=270,width=8.cm]{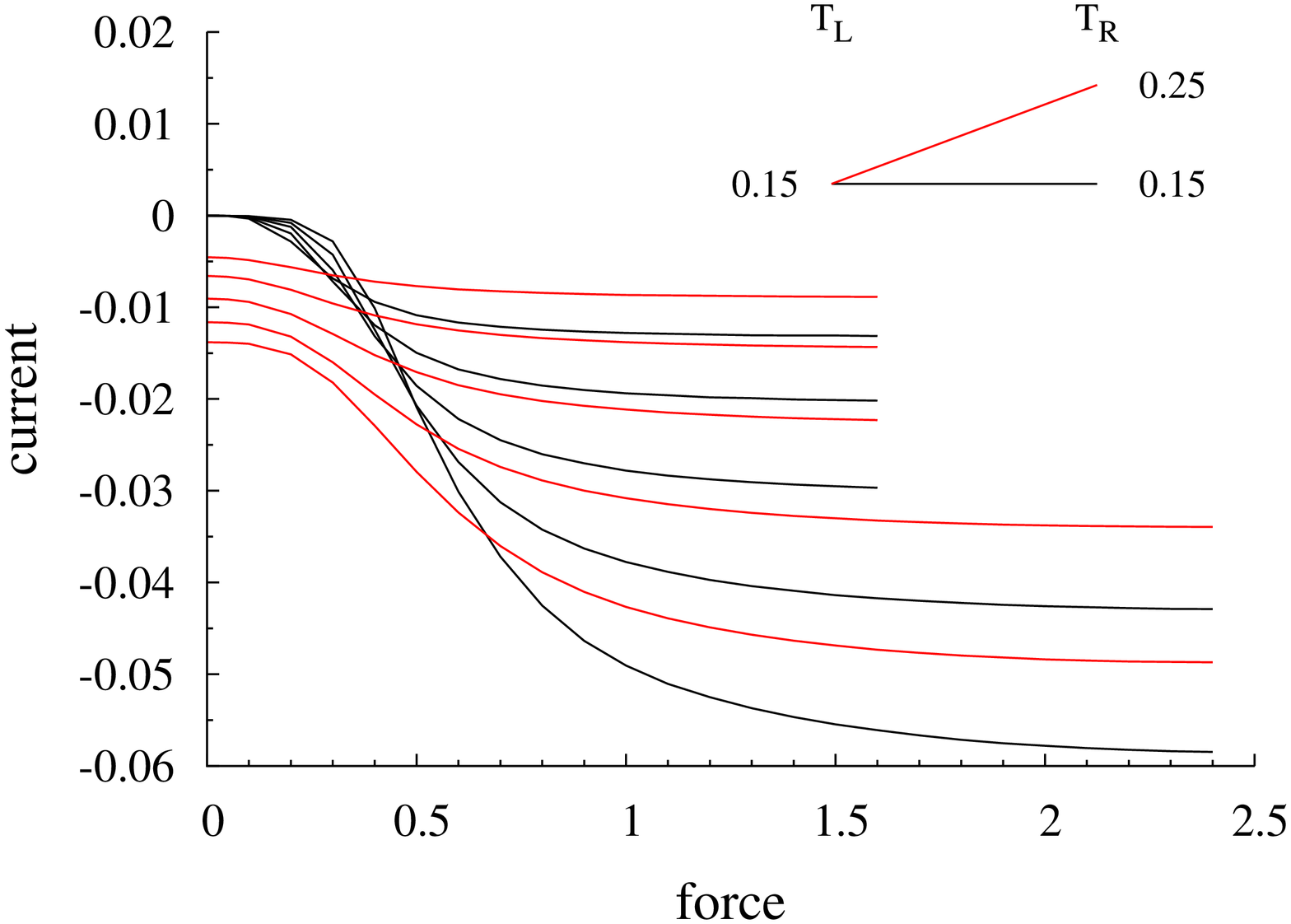}
\caption{\label{fig:TLfixed}
Comparison of the currents for a fixed temperature at the left end and various
temperature differences.
From top to bottom: decreasing system sizes $N=2048,1024,512,256,128$
(the ordering is the same for all situations considered).
}
\end{center}
\end{figure}

We consider the following situations: 
\begin{enumerate}[(i)]
\item same temperatures on the left and on the right: 
$(T_{\rm L},T_{\rm R}) = (0.20,0.20)$ or $(0.15,0.15)$;
\item hot left end and cold right end: 
$(T_{\rm L},T_{\rm R}) = (0.25,0.15)$, $(0.20,0.15)$ or $(0.25,0.20)$;
\item cold left end and hot right end: $(T_{\rm L},T_{\rm R}) = (0.15,0.25)$.
\end{enumerate}
Currents are computed as a function of the magnitude $F$ of the 
non-gradient forcing term for systems of different lengths: 
$N = 128$, 256, 512, 1024, 2048.
Recall that local equilibrium holds at the leading order, so that the
energy current is induced by the first order corrections in~$1/N$. 

We first compare the currents when the temperature on the right 
end is fixed  (see Figure~\ref{fig:TRfixed}).
As expected, the negative current induced by the mechanical forcing is reduced by
the opposite, positive thermal current.

We next compare the currents at fixed temperature difference $T_{\rm R}
- T_{\rm L}$, for different average temperatures (see Figure~\ref{fig:dTfixed}).
In this case, we observe that, for strong mechanical forcings, 
the current is enhanced when the average temperature decreases, while
the opposite happens when the mechanical forcing is small.

We finally turn to the most interesting situation. The temperature 
on the left end is fixed 
and the temperature on the right end varies (see Figure~\ref{fig:TLfixed}).
In this case, counterintuitive results are observed for large mechanical forcings:
The total current is enhanced as $T_{\rm R}$ decreases, even though, in such a situation,
the thermal gradient is in the opposite direction. The mechanical
forcing induces a negative current, while the thermal gradient induces
(in the absence of any force) a positive current. The combined effect
of both mechanical and thermal forcings induces a negative current {\em larger}
(in absolute value) than the one in the absence of any thermal
gradient!

\section{Discussion of the results}

In conclusion, for large mechanical forcings $F$, we observe that
\begin{enumerate}[(a)]
\item when $T_{\rm R}$ is fixed, the current varies qualitatively as 
when there is no mechanical forcing: The absolute value of the current
increases when $T_{\rm L}$ decreases, which means that the current induced by the thermal forcing
and the current induced by the mechanical forcing are somewhat additive. In this case, a 
positive thermal conductivity is observed (for a fixed value $F$ of the mechanical forcing, 
considering only the response in the limit when $T_{\rm R} - T_{\rm L} \to 0$).
\item when $T_{\rm L}$ is fixed, the current has a surprising behavior:
  Its absolute value increases when $T_{\rm R}$
  decreases. This means that the thermal
  forcing, which is naively expected to reduce the current induced by
  the mechanical forcing, actually enhances it. In this case, a negative
  thermal conductivity is observed (again, for a fixed value $F$ of the
  mechanical forcing). 
\end{enumerate}
A possible interpretation is based on the fact that, for such a system,
the thermal conductivity is a decreasing function of the
temperature when $F$ is large (see Figure~\ref{fig:dTfixed}). It is
possible that, by lowering $T_{\rm R}$ and thus 
increasing the conductivity at the right end, one makes the system more
sensitive to the mechanical forcing. 
The increased mechanical current
may hence counterbalance the increased opposite thermal current.

An interesting question which we did not discuss here 
is the scaling of the energy current as a function of the system
size when $F \neq 0$. Some preliminary results suggest that the thermal
conductivity is finite, as when
$F=0$, but this question definitely calls for additional 
studies.

\acknowledgments{
  This work is supported in part by the Agence Nationale de la Recherche, 
  under grants ANR-09-BLAN-0216-01 (MEGAS), ANR-10-BLAN 0108 (SHEPI)
  and by the European Advanced Grant \emph{Macroscopic Laws and
    Dynamical Systems} (MALADY) (ERC AdG 246953).
}


\end{document}